\documentclass[prd, aps, superscriptaddress, preprintnumbers, twocolumn, floatfix, nofootinbib]{revtex4}
\pdfoutput=1

\usepackage{amsfonts}
\usepackage{amsmath}
\usepackage{amssymb}
\usepackage{bm}
\usepackage{dcolumn}
\usepackage{graphicx}   
\usepackage[latin1]{inputenc}
\usepackage{latexsym}
\usepackage{rotating}
\usepackage{hyperref}
\usepackage{graphicx}
\usepackage{color}

\newcommand\be{\begin{equation}}
\newcommand\ba{\begin{eqnarray}}
\newcommand\ee{\end{equation}}
\newcommand\ea{\end{eqnarray}}

\begin{document}

\title {Cosmic Rays and Spectral Distortions from Collapsing Textures}

\author{Robert Brandenberger}
\email{rhb@physics.mcgill.ca}
\affiliation{Department of Physics, McGill University, Montr\'{e}al, QC, H3A 2T8, Canada} 

\author{Bryce Cyr}
\email{bryce.cyr@mail.mcgill.ca}
\affiliation{Department of Physics, McGill University, Montr\'{e}al, QC, H3A 2T8, Canada} 

\author{Hao Jiao}
\email{jiaohao@mail.ustc.edu.cn}
\affiliation{CAS Key Laboratory for Research in Galaxies and Cosmology, Department of Astronomy,
University of Science and Technology of China, Chinese Academy of Sciences, Hefei, Anhui 230026, China}

\date{\today}

\begin{abstract}

We compute the energy spectrum of photons and neutrinos produced by the unwinding of a scaling distribution of cosmic textures, and discuss the implications for the spectrum of high energy cosmic rays, and for CMB spectral distortions. Textures lead to a contribution to the photon flux which scales as $E^3 F(E) \sim E^{3/2}$. Hence, the tightest constraints on the texture model come from the highest energies from which primordial photons can reach us without being scattered by the CMB and other foregrounds. Textures lead to both $\mu$ type and $y$ type distortions. While the constraints on the texture model coming from the current COBE bounds are weaker than the bounds from the angular power spectrum of the CMB, future surveys such as PIXIE can lead to stronger bounds. The high energy neutrino flux is constrained by data from the Pierre Auger experiment and yields a bound on the energy scale of textures which is competitive with CMB bounds.

\end{abstract}

\maketitle

\section{Introduction} 
\label{sec:intro}

It is of great interest to study the cosmological consequences of particle physics models beyond the Standard Model which have defect solutions. The defects which inevitably form in the early universe in such models persist to the present time and lead to signatures in cosmological surveys. Thus, the study of topological defects in cosmology leads to a close connection between particle physics and cosmology (see e.g. \cite{RHBrev2} for a discussion). If evidence for defects is found in observations, we would learn which class of particle physics models beyond the Standard Model is favored. If the predicted signatures are not found, we can constrain the energy scale of particle physics models which predict the signatures. Particle physics models which predict domain wall solutions are already ruled out by cosmology \cite{DW}, since they would overclose the universe. Models which predict monopoles are also constrained \cite{MP}. There has been a lot of work on the cosmological signatures of cosmic strings (see e.g. \cite{VS, HK, RHBrev}). Here, we will focus on another class of defects, namely global textures.

Low energy effective field theories in which a global symmetry is spontaneously broken such that the vacuum manifold ${\cal{M}}$ has a nontrivial third homotopy group $\Pi_3({\cal{M}})$ admit defect solutions called {\it global textures} \cite{Texture} (see e.g. \cite{RHBrev} for reviews of the cosmology of topological defects, including discussions of textures). Like global monopoles, textures are extended configurations of field energy with the topology of a sphere. By causality \cite{Kibble}, textures will form during the symmetry breaking phase transition in the early universe, and their initial radial extent will be comparable to the Hubble radius. The same causality argument implies that at all late times there will be of the order one texture per Hubble volume. Hence, textures can leave potentially observable imprints even at late times.

In contrast to global string and global monopole defects, textures are not stable. A texture with initial size comparable to the Hubble radius at time $t$ will collapse at close to the speed of light and ``unwind''. In this process, a fraction of the initial energy stored in the texture configuration will be released as particles. An order one fraction of the released energy ends up in the form of photons, including high energy photons and ultra-high energy neutrinos. Photons emitted in the redshift interval between $z = 10^6$ and the redshift of recombination will lead to spectral distortions of the Cosmic Microwave Background (CMB). In this paper, we compute the spectrum of high energy gamma rays produced by texture collapse, and we estimate the magnitude of the induced spectral distortions.

An example of a low energy theory which has texture solutions is quantum chromodynamics (QCD). In the limit of massless quarks, the theory has a global $SU(3) \times SU(3)$ symmetry which gets spontaneously broken to the diagonal subgroup $SU(3)$, in the process generating an octet of pseudoscalar mesons which are the Goldstone bosons of the symmetry breaking. In theories beyond the Standard Model of particle physics there may be similar global symmetries which are spontaneously broken at an energy scale $\eta$, an energy scale of new physics. Since texture configurations contain trapped energy, they can contribute to structure formation in the early universe \cite{Texture, Gooding} and induce cosmic microwave background (CMB) anisotropies. The induced CMB anisotropies have an angular power spectrum which do not have \cite{Pen} the acoustinc oscillations which have been observed. Hence, observations of CMB anisotropies lead to an upper bound on the energy scale $\eta$ of new physics with texture solutions \cite{Durrer}. Textures with an energy scale slightly lower than the upper bound may, however, help explain the observed cold spot in the CMB maps \cite{cold}. Since textures form nonlinear density perturbations at arbitarily early times, they may contribute to the seeds for the observed super-massive black holes at high redshifts (in the same way that cosmic string loops can \cite{CSBH}). This was discussed in a recent paper \cite{us} by two of us. In fact, there is a range of values of $\eta$ where the number density of nonlinear seed fluctuations at high redshifts is larger than that produced in the standard $\Lambda$CDM paradigm of early universe cosmology.

Whereas most work done on the cosmological implications of textures has focused on their gravitational effects, here we will study effects on the electromagnetic spectrum. We compute the spectrum of induced high energy cosmic rays and compare our results with the observational bounds, and we estimate the amplitude of the induced spectral distortions of the CMB.

We begin this article with a brief review of textures, focusing on the scaling solution which is achieved. In Section 3 we then compute the spectrum of photons and neutrinos emitted from  textures collapsing. In the case of photons, the relevant collapse time begins at the time of recombination. As neutrino decoupling takes place long
before recombination, collapsing textures in the primordial plasma will also
contribute to their production. In Section 4 we compute the total energy released by textures collapsing during the time interval when the emitted photons can produce spectral distortions. We conclude with a discussion section.

We use natural units in which the speed of light, Planck's constant and Boltzmann's constant are all set to $1$. We work in the context of a Friedmann-Lemaitre-Robertson-Walker metric
\be
ds^2 \, = \, dt^2 - a(t)^2 d{\bf x}^2 \, ,
\ee
where $t$ is physical time, ${\bf x}$ are comoving spatial coordinates (we assume that the spatial curvature vanishes), and $a(t)$ is the scale factor in terms of which the Hubble parameter is defined as
\be
H(t) \, = \, \frac{\dot{a}}{a} \, ,
\ee
and its inverse is the Hubble radius. We often work in terms of the cosmological redshift $z(t)$ defined by
\be
\frac{a(t_0)}{a(t)} \, \equiv \, z(t) + 1 \, .
\ee

\section{Review of Global Textures and their Scaling Solution} \label{review}

To get a feeling for what a global texture is, consider a scalar field theory with four real scalar fields $\phi_i$ and a symmetry breaking potential of the standard form
\be
V(\phi) \, = \, \frac{\lambda}{4} \biggl( \sum_{i = 1}^4 \phi_i^2 - \eta^2   \biggr)^2 \, ,
\ee
where $\eta$ is the symmetry breaking scale and $\lambda$ is a dimensionless coupling constant. At high temperatures, finite temperature corrections to the potential force $\phi(x) = 0$ at all points in space. Below the critical temperature which is of the order $\eta$ (modulo coupling constants) the confining force disappears and $\phi(x)$ rolls down to a minimum of its potential. The space of field values which minimize the potential energy is called the vacuum manifold ${\cal{M}}$ and is given by
\be
\mathcal{M} \, = \left[ \phi \,\, \bigg{|} \, \,  \sum_{i = 1}^4 \phi_i^2 = \eta^2  \right]
\ee
which in our case is a three sphere $S^3$. It has the property that the third homotopy group is nontrivial
\be \label{crit}
\Pi_3({\cal{M}}) \, \neq \, 1 \, .
\ee
All field theories with the property (\ref{crit}) have texture solutions.

By causality \cite{Kibble} there can be no correlation on scales larger than the Hubble radius. Hence, there is a finite probability $c$ of the order one that in any Hubble volume the field $\phi(x)$ covers the entire vacuum manifold \footnote{The study of \cite{Tomislav} shows that the probability is approximately $c \simeq 0.04$.}. Such a configuration winds the entire vacuum manifold.

For our Lagrangian, a texture configuration with winding number $1$ and with spherically symmetric energy distribution is
\be \label{textconfig}
\phi(t,x, y, z) \, = \, 
\eta \bigl( {\rm{cos}} \chi(r),\frac{x}{r} {\rm{sin}} \chi(r) ,\frac{y}{r} {\rm{sin}} \chi(r), \frac{z}{r} {\rm{sin}} \chi(r) \bigr)
  \, 
\ee
where $r^2 = x^2 + y^2 + z^2$. The radial function vanishes at the origin (i.e. $\chi(0) =  0$) and tends to the value $\pi$ as $r \rightarrow \infty$. The {\it effective radius} of the texture can be defined to be the radius where $\chi(r_c) = \pi/2$, i.e. the radius $r_c$ where the texture configuration (restricted to the sphere of radius $r_c$ in space) wraps the equator sphere of the vacuum manifold ${\cal{M}}$.

Note that, unlike for monopole, cosmic string and domain wall defects, in the case of textures it is possible for the scalar field to be everywhere in the vacuum manifold. Hence, there is no trapped potential energy, but only gradient energy. In the case of theories with a local symmetry, the gradient energy in the scalar field can be locally compensated by the gauge fields, and hence there are no well-localized local texture configurations.

The total gradient energy in the texture configuration (\ref{textconfig}) can decrease if $r_c$ decreases. Since we are dealing with relativistic field theories, the speed of contraction is of the order of the speed of light. The energy density of the collapsing texture configuration becomes \cite{TuSp}
\be \label{grad}
\rho(r, s) \, \simeq \, 2 \frac{r^2 + 3 s^2}{(r^2 + s^2)^2} \eta^2
\ee
where $s$ is the shifted time such that the collapse happens at $s = 0$. This formula is a good approximation up the radius $r_c$. 

As $s \rightarrow 0$ the gradient energy density (\ref{grad}) of the collapsing texture diverges at the origin. When the energy density at the origin becomes comparable to the potential energy, it becomes favourable for the texture configuration to ``unwind'', i.e. $\phi(x)$ in a region close to the center leaves the vacuum manifold and $\chi$ jumps from $\chi \sim 0$ to $\chi \sim \pi$. The time when this happens is \cite{us}
\be
s_{uw} \, \sim \, \sqrt{6 \lambda^{-1}} \eta^{-1} \, ,
\ee
and the radius of the central region where the unwinding happens is
\be
r_{uw} \, \sim \, \lambda^{-1/2} \eta^{-1} \, .
\ee
After the unwinding, the field configuration is no longer topologically confined, and will radiate outwards. The total energy in a texture is obtained by integrating (\ref{grad}) from $r = 0$ to $r = r_c$, setting $s = r_c$ (the collapse time). The result is
\be
E \, \sim \, 8 \pi \eta^2 r_c \, .
\ee
For a texture becoming dynamical at the time $t_f$ we can set $r_c \simeq t_f / 2$, and hence the amount of energy released by a texture forming at time $t_f$ is
\be \label{energy}
E(t_f) \, \sim \, 4 \pi \eta^2 t_f \, .
\ee

Textures first form during the symmetry breaking phase transition. For any sphere in space with radius $R$, there is a probability $c$ of the order one that the configuration of $\phi$ in that sphere will wind the vacuum manifold and that hence a texture will form. Textures with radius smaller than the Hubble radius, i.e. $R < t$ will collapse. However, by causality the field configuration on larger scales is still random. Hence, at all times $t$, there will be a probability $c$ that a texture of size $r_c \sim t$ will form and start to contract. This process of texture configurations entering the Hubble radius, becoming dynamical and starting to collapse will continue to the present time. This is the {\it scaling solution} for textures: at any time $t$ after the phase transition, there will be $c$ textures per Hubble radius of roughly Hubble size which begin to contract and unwind within a Hubble time. In the following we will compute the spectrum of cosmic rays produced by this scaling solution of textures.

\section{Spectrum of Cosmic Rays Produced by Texture Collapse} \label{analysis}

The primary particles which result from the texture unwinding are $\phi$ quanta. The particles associated with $\phi$ are, however, unstable, and their decay will generate a jet of Standard Model elementary particles, in particular photons. This process is similar to the production of a jet of low energy particles in laboratory decays of unstable particles, or from the annihilation of a cosmic string cusp \cite{cuspannih}. In the case of cosmic string cusp decay, the spectrum of resulting photons and neutrinos was computed in \cite{Jane1} and \cite{Jane2}, respectively. Here we will perform a similar calculation for the photon and neutrino fluxes resulting from texture decay.

To obtain the spectra of photons and neutrinos from a texture collapse we assume that the primary particles released (the $\phi$ particles) decay into jets of Standard Model quarks and leptons, and that these then produce photons and neutrinos in agreement with the QCD multiplicity functions. Following \cite{HSW}, we take the spectra at energies $E < \eta$ from a texture collapsing at time $t_f$ to be given by
\be \label{number}
\frac{d N}{d E^{\prime}}( E^{\prime}, t_f) \, = \, \frac{f_i}{2} E(t_f) \bigl( \frac{\eta}{E^{\prime}} \bigr)^{3/2}
\frac{1}{\eta^2} \, ,
\ee
where $N( E^{\prime}, t_f)$ is the number per unit energy interval of photons or neutrinos of energy $E^{\prime}$ at time $t_f$, and $f_i$ is the fraction of the energy of texture annihilation which goes into photons $f_i = f_e$ and neutrinos $f_i = f_{\nu}$, respectively. The above fragmentation functions hold for energies larger than the mass of the $\pi_0$ meson.

Let us first consider the photon spectrum.
If we want the energy density $d\rho / dE$ for unit interval of energy at energy $E$ and time $t$, then we have to integrate over all texture formation times $t_f$. For each value of $t_f$, we need to consider the number of photons with energy $E^{\prime}$ at time $t_f$ which redshifts to $E$ at time $t$. We must finally take into account the redshifting of the number density of the produced photons. In this way, we get
\be \label{fracen1}
\frac{d\rho}{dE} (E, t) \, = \, E \int_{t_i(E, t)}^{t} dt_f \frac{dN}{dE^{\prime}} (E^{\prime}, t_f) c t_f^{-4} 
\frac{dE^{\prime}}{dE} \bigl( \frac{z(t)}{z(t_f)} \bigr)^3 \, ,
\ee
where we have used the fact that there are $c$ textures for Hubble volume per Hubble time (thus giving the factors $c t_f^{-4}$ in the above, and the Jacobian of the transformation between $E^{\prime}$ and $E$. Note that in the above
\be
E^{\prime} \, = \, \frac{z(t_f)}{z(t)} E \, .
\ee
The time $t_i(E, t)$ is the larger of the following two times: the time $t_c$ of the phase transition, and the time $t_b(E, t)$ before which all photons emitted by a texture have redshifted energy less than $E$ by the time $t$. This time is given by
\be
\frac{z(t_b(E, t))}{z(t)} \, = \, \frac{\eta}{E} \, .
\ee
Inserting (\ref{number}) and (\ref{energy}) into (\ref{fracen1}) we obtain
\be
\frac{d\rho}{dE} (E, t) \, \sim \,  2 \pi c f_e \eta^{3/2}  E^{-1/2} \int_{t_i(E, t)}^{t} dt_f t_f^{-3} 
\bigl( \frac{z(t)}{z(t_f)} \bigr)^{7/2} \, .
\ee
Evaluating the result for times before the time $t_{eq}$ of equal matter and radiation we obtain
\be
\frac{d\rho}{dE} (E, t) \, \sim \,  8 \pi c f_e \eta^{3/2}  E^{-1/2} t^{-2} 
\bigl[  \bigl( \frac{t}{t_i(E, t)} \bigr)^{1/4} - 1 \bigr] \, ,
\ee
the time integral being dominated by the lower integration end. 

For $t > t_{eq}$ we divide the integration over $t_f$ into the interval before $t_{eq}$  and the interval after $t_{eq}$. The contribution of textures at $t_f > t_{eq}$ gives
\be \label{result}
\frac{d\rho}{dE} (E, t) \, \sim \,  6 \pi c f_e \eta^{3/2}  E^{-1/2} t^{-2} \bigl[ 1 -  \bigl( \frac{t_{eq}}{t} \bigr)^{1/3} \bigr] \, ,
\ee
assuming that $t_i(E, t) < t_{eq}$. If $t_i(E, t) > t_{eq}$, then the factor $t_{eq}$ in the above is replaced by $t_i(E, t)$.

The contribution from textures formed before $t_{eq}$ gives
\ba \label{early}
\frac{d\rho}{dE} (E, t) \, &\sim& \,  8 \pi c f_e \eta^{3/2}  E^{-1/2} t^{-2} \\
& & \, \bigl( \frac{t_{eq}}{t_i(E, t)} \bigr)^{1/4}
\bigl( \frac{t_{eq}}{t} \bigr)^{1/3} \bigl[ 1 - \bigl( \frac{t_i(E, t)}{t_{eq}} \bigr)^{1/4}  \bigr]\, . \nonumber
\ea
If we are interested in the spectrum of high energy cosmic rays today, then photons released from textures before $t_{eq}$ do not contribute. Those released before a redshift of $z \sim 10^6$ will thermalize, and those created between that redshift and recombination will scatter off the plasma and lead to spectral distortions of the CMB (as studied in the next section), but they will not survive as primary cosmic rays. Hence, in the following we will only consider the contribution (\ref{result}).

We now compare the predicted cosmic ray flux from textures with the observational upper bounds. These bounds are usually expressed in terms of the flux $F(E)$, where
\be
F(E) \, = \, E^{-1} \frac{d\rho}{dE} \, .
\ee
Since the measured flux and the upper flux limits decay rapidly as a function of $E$, the quantity which is usually used is $E^3 F(E)$. From (\ref{result}) we see that the texture model predicts
\be \label{prediction}
E^3 F(E) \, \sim \, 6 \pi c f_e (E \eta)^{3/2} t^{-2} \, ,
\ee
which is a rapidly increasing function of $E$. Hence, the tightest bounds will come from the largest values of $E$. There is a caveat, however: as discussed in detail in \cite{Jane3}, photons energies greater than about $10^2 {\rm{GeV}}$ cannot travel over cosmological distances. They scatter off the CMB and off the infrared and optical backgrounds. Hence, the strongest (but conservative) bound on the symmetry breaking scale $\eta$ will come from comparing our predictions with the observational limits at $E = 10^2 {\rm{GeV}}$ which are \cite{Jane3,Swordy,Angelo} (in the usual units used in cosmic ray physics)
\be \label{bound}
E^3 F(E) \, < \, 10^{21} b {\rm{eV}}^2 {\rm{m}}^{-2} {\rm{sec}}^{-1} {\rm{sr}}^{-1} \, ,
\ee
where we have introduced a factor $b$ to take into account future improvements of the bound (at the moment the bound is $b = 1$). Comparing the prediction (\ref{prediction}) with the observational bound (\ref{bound}) we obtain
\be
G \eta^2 \, < \bigl( \frac{1}{6 \pi c f_e} \bigr)^{4/3} 10^{7/3} b^{4/3} \bigl( \frac{E_2}{E} \bigr)^2 \, ,
\ee
where $E_2 = 10^2 {\rm{GeV}}$. At the present time, using $b = 1, f_e \sim 1$ and $E = E_2$, there is no bound on the symmetry breaking scale $\eta$ (since $G \eta^2$ cannot be larger than unity). This result contrasts with actual bounds \cite{Durrer} which come from the effects of textures on the angular power spectrum of the CMB. If scattering off of the infrared and optical backgrounds is neglected, then the photon spectra could be extrapolated up to $E = 10^5 {\rm{GeV}}$. In this case, for $f \sim 1$ there would be a bound of the form  (with $b = 1$)
\be
G \eta^2 \, < \, 10^{-4} \, ,
\ee
which is still a few orders of magnitude weaker than the bound coming from the CMB.

The calculation of the high energy neutrino flux is completely analogous to the computation of the photon flux after replacing $f_e$ by $f_{\nu}$. However, neutrinos can reach us from much earlier times. For energies $E < z_{eq}^{-1} \eta$, the earliest textures producing neutrinos which today have energy $E$ were formed at a time $t_i(E)$ given by
\be
t_i(E) \, = t_{eq} z_{eq}^2 \bigl( \frac{E}{\eta} \bigr)^2 \, .
\ee
Inserting this result into (\ref{early}) yields
\be \label{nflux}
E^3 F(E) \simeq \, 8 \pi f_{\nu} c E \eta^2 z_{eq}^{-1} t_0^{-2} \, .
\ee

We can derive an upper bound on $G\eta^2$ by demanding that the flux (\ref{nflux}) be smaller than the upper bound which follows from the Pierre Auger experiment. This experiment detected no ultra-high energy neutrino events in the accessible energy range $10^8 {\rm GeV} < E < 2.5 \times 10^{10} {\rm GeV}$ and obtained the bound \cite{PA}
\be \label{nbound}
E^3 F(E) \, < \, 6.4 \times 10^{21} E_8 {\rm ev}^2 {\rm m}^{-2} {\rm s}^{-1} {\rm sr}^{-1} \, ,
\ee
where $E_8$ is the value of the energy in units of $10^8 {\rm GeV}$. Comparing this bound with the predicted flux from (\ref{nflux}) yields the constraint
\be
G\eta^2 \, < \, \frac{1}{f_{\nu}} 10^{-9} \, ,
\ee
which is in fact slightly stronger than the bound from the CMB, as long as the neutrino fragmentation fraction $f_{\nu}$ is not much smaller than $1$. Note that the range of values of $\eta$ which are ruled out by the above constraint are in the range for which $E < z_{eq}^{-1} \eta$ is satisfied for the value of the energy $E = 10^8 {\rm GeV}$, a condition which was assumed in the calculation. This is a self-consistency check of our analysis.

To conclude our discussion of limits from ultra-high energy neutrinos, we must check that the neutrinos indeed can travel freely until the present time beginning at the redshift $z(t_i(E))$. As discussed in \cite{Jane2}, the neutrino free streaming redshift $z_{IA}(E)$ below which neutrinos of energy $E$ can freely stream to us is given by
\be
z_{IA} \, \sim \, 3 \times 10^{10} \bigl( \frac{E}{1 {\rm GeV}} \bigr)^{0.35} \, .
\ee
Evaluated at the energy $E = 10^8 {\rm GeV}$ we find that neutrino interactions are indeed negligible (i.e. $z_{IA} > z(t_i(E))$) if $\eta$ is smaller than about $10^{16} {\rm GeV}$. Larger values of $\eta$ are not of great interest to us since textures with such a scale are already ruled out by the CMB data.

\section{Induced Spectral Distortions}

Photons emitted in the redshift range between $z_1 \simeq 3 \times 10^6$ and the time of recombination cannot fully thermalize and give rise to CMB spectral distortions\footnote{The effect is similar to the one produced by cusp annihilation from cosmic string loops which has been studied in \cite{Maddy}.}. Roughly speaking, those emitted between $z_1$ and $z_2 \sim 10^5$ yield  $\mu$ distortions, those emitted between $z_2$ and recombination give rise to $y$ distortions (see e.g. \cite{Chluba} \cite{Tashiro} for reviews).

A $\mu$ distortion is sourced when energy release occurs in a redshift window where photon number changing interactions (Bremsstrahlung and double Compton scattering) are frozen out, and only a kinetic equilibrium can be reached in the photon plasma. The spectrum then acquires a Bose-Einstein like correction, where the $\mu$ parameter is given by the expression
\be \label{mu}
\mu \, \equiv \, -\frac{1}{0.7} \int_{z_{\rm{C}}}^{\infty} dz \frac{1}{\rho_{\gamma}} \frac{dU}{dz} \cdot e^{-(z/z_{\rm{DC}})^{5/2}},
\ee
Where $\rho_{\gamma}$ is the usual photon energy density at a redshift $z$, while $z_{\rm{DC}} \approx 2 \times 10^6$ is the approximate redshift at which double Compton scattering freezes out for the usual values of Helium abundance and matter energy density. $z_{\rm{C}} \approx 10^5$ is the redshift that Compton scattering freezes out, marking an end to the $\mu$ era. Finally, $dU/dz$ is the rate of energy density injection from the decay of cosmic textures, given by
\be \label{releaseRate}
\frac{dU}{dz} = -8\pi c f_e \frac{\eta^2}{t_0^2} z^3
\ee
This is computed by taking the number density of textures per unit time and multiplying by their individual energy release, so at the time of formation, $t_f$, we have $dU/dt_f = c t_f^{-4}\cdot E(t_f)$. We also consider the fact that textures produced in the $\mu$ distortion window will collapse within a Hubble time, releasing a fraction, $f$, of their energy into photons. The exponential in (\ref{mu}) acts to damp out contributions for $z > z_{\rm{DC}}$, so with this in mind we can find that the induced $\mu$ distortion is
\be \label{muResult}
\mu \approx \frac{8\pi}{0.7} c f_e (G \eta^2) \ln\left(\frac{z_{\rm{DC}}}{z_{\rm{C}}} \right) \sim 6 f (G\eta^2)
\ee
Where we have made use of $c \approx 0.04$, and noting that the CMB energy density at redshift $z$ is $\rho_{\gamma} \approx (3/4) G^{-1} t_0^{-2} z^4$.

A primordial\footnote{y-distortions are also sourced after recombination, for example through the upscattering of CMB photons as they pass through hot clumps of electrons in our line-of-sight} y-distortion is induced in the CMB spectrum for energy releases in the redshift range between recombination and $z_{\rm{C}}$, as we are no longer able to even sustain a kinetic equilibrium in the photon plasma. This distortion is primarily sourced by the fact that external energy injection heats up the pre-recombination electrons, which transfer their non-thermal energy to the rest of the photons. Traditionally, the y parameter is written as
\be 
y = \int_{t_{\rm{C}}}^{t_{\rm{rec}}} dt \frac{T_e - T}{m_e} n_e \sigma_T
\ee
where $T_e$ is the electron temperature, $T$ is the CMB photon temperature, $n_e$ is the number density of free electrons at time $t$, and $\sigma_T$ is the Thomson scattering cross section. It is manifest from this expression that a decoupling between the electron and photon temperatures will induce a y-distortion. Under reasonable assumptions (neglecting cosmic expansion when compared to the timescale of Compton scattering, and assuming the injected photons perturb the electron temperature only slightly), one can recast this expression in a more applicable way to energy release scenarios
\be
y \approx -\frac{1}{4} \int_{z_{\rm{rec}}}^{z_{\rm{C}}} dz \frac{1}{\rho_{\gamma}} \frac{dU}{dz}
\ee
Since the energy release mechanism we have is insensitive to the transition between the y and $\mu$ eras, we can simply use (\ref{releaseRate}) once again to compute our contribution to the y distortion. Doing this yields 
\be \label{yResult}
y \approx \frac{8\pi}{3} c f_e (G \eta^2) \ln \left(\frac{z_{\rm{C}}}{z_{\rm{rec}}} \right) \sim 1.5 f (G \eta^2)
\ee
At the present time, the best bounds on $\mu$ and y are from the COBE satellite
\be
\mu \, < \, 9 \times 10^{-5} \, , \,\,\, {\rm{COBE}}
\ee
\be
y \, < \,  10^{-5} \, , \,\,\,\,\,\,\,\,\,\,\,\,\,\, {\rm{COBE}}
\ee
Of these two bounds, the y distortion yields the tightest constraint on the parameter space
\be \label{bound1}
f_e (G \eta^2) \, < \, 6.7 \times 10^{-6} \, .
\ee
Given that $f_e < 1$, this bound is slightly weaker than the constraint which comes from the CMB angular power spectrum. The proposed PIXIE experiment will have a sensitivity of \cite{PIXIE}
\be
\mu \, < \, 10^{-8} \, , \,\,\, {\rm{PIXIE}} \, ,
\ee
\be
y \, < \, 2\times 10^{-9} \, , \,\,\, {\rm{PIXIE}} \, ,
\ee
and with this experiment the bound on the scale $\eta$ could be strengthened to read (this time from the proposed sensitivity on the $\mu$ parameter)
\be \label{bound2}
f_e (G \eta^2) \, < \, 1.7 \times 10^{-9} \, ,
\ee
If no distortions are seen. This bound would be significantly stronger than the current bounds from the CMB. Since these textures would decay through the entire ($\mu$ and y) distortion window, they could be probed even more precisely as decays around $z_{\rm{C}}$ may encode information on the exact time dependence of the energy release. A lack of a detection of this form would put serious constraints on symmetry breaking patterns that would give rise to textures in the early universe.

Note that due to their feeble interactions with the primordial plasma, neutrinos from texture collapse do not lead to any spectral distortions.

\section{Conclusions and Discussion} \label{conclusion}

We have studied the emission of photons and neutrinos from a network of unwinding cosmic textures. We computed the spectrum of high energy cosmic rays and compared the results with the current observational bounds. For photons, the bounds satisfied for all values of the energy scale $\eta$ of the textures which are compatible with the constraints from the angular power spectrum of the CMB. In the case of neutrinos, we obtain an upper bound on the symmetry breaking scale $\eta$ which is comparable (and possibly slightly stronger) than the bounds from the CMB angular power spectrum. We then computed the induced spectral distortions of the CMB produced by textures unwinding in the redshift interval between $z = 10^6$ and recombination. Whereas the current bounds from the COBE experiment are not competitive with the CMB angular power spectrum constraint, future missions such as PIXIE will be able to yield stronger constraints on $\eta$.

The photon flux from unwinding textures can effect more cosmological observables. For example, the extra photon radiation at radio wavelengths may lead to an absorption feature \cite{Bowman} in the global 21cm signal, in the same way that decaying cosmic string loops may yield such a signal \cite{CS21cm}. It is unlikely, however, that unwinding events can explain the origins of fast radio bursts (FRBs), as the decay pattern to low energy photons suffers from similar problems to those photons produced off of cosmic string cusps \cite{FRB}. On top of that, the number density of these undwinding events is much smaller than that of string loops studied in \cite{FRB}, and so this model would also suffer from being unable to explain the current number densities of FRBs.

Global monopoles \cite{GM} are monopole defects in a theory with a global symmetry. The number density and topology of these defects is similar to that of textures.  Note that for a global monopole, the total energy of a monopoe is also comparable to that of a texture because the energy is dominated by the long range gradients, as it is for global textures. Whereas global monopoles are stable, unlike textures, monopole-antimonopole annihilation will lead to a scaling solution of defects very similar to that of textures, except that the constant $c$ describing the number density of these defects, will be different ($c \sim 1.2$ \cite{GM2}). The spectrum of cosmic rays and the induced CMB spectral distortions will hence be analogous to the ones studied here. Since the value of $c$ is larger by a factor of about $30$, the resulting bounds on $G\eta^2$ will be tighter by a similar factor.

\section*{Acknowledgement}
\noindent RB wishes to thank the Pauli Center and the Institutes of Theoretical Physics and of Particle- and Astrophysics of the ETH for hospitality. The research of RB at McGill is supported in part by funds from NSERC and from the Canada Research Chair program. BC thanks support from a Vanier-CGS fellowship and from NSERC. HJ is supported in part by the Undergraduate Education Office of USTC. She also wishes to thank Professor Lavinia Heisenberg for an invitation to the ETH.  We wish to thank an anonymous referee for suggesting we consider not only the photon but also the neutrino spectrum.

\end{document}